# A Vertically Resolved MSE Framework Highlights the Role of the Boundary Layer in Convective Self-Aggregation

Lin Yao[1], Da Yang*[1,2], and Zhe-Min Tan[3]

[1]University of California, Davis, CA, USA.

[2]Lawrance Berkeley National Laboratory, Berkeley, CA, USA.

[3]Nanjing University, Nanjing, China.

*Corresponding author: Da Yang, dayang@ucdavis.edu

**Abstract**

Convective self-aggregation refers to a phenomenon in which random convection can self-organize into large-scale clusters over an ocean surface with uniform temperature in cloud-resolving models. Previous literature studies convective aggregation primarily by analyzing vertically integrated (VI) moist static energy (MSE) variance. That is the global MSE variance, including both the local MSE variance at a given altitude and the covariance of MSE anomalies between different altitudes. Here we present a vertically resolved (VR) MSE framework that focuses on the local MSE variance to study convective self-aggregation. Using a cloud-resolving simulation, we show that the development of self-aggregation is associated with an increase of local MSE variance, and that the diabatic and adiabatic generation of the MSE variance is mainly dominated by the boundary layer (BL). The results agree with recent numerical simulation results and the available potential energy analyses showing that the BL plays a key role in the development of self-aggregation. We further present a detailed comparison between the global and local MSE variance frameworks in their mathematical formulation and diagnostic results, highlighting their differences.

# 1. Introduction

Uniformly distributed convection can spontaneously cluster into large-scale upwelling areas over an ocean surface with uniform temperature in cloud-resolving model (CRM) simulations. This phenomenon is known as convective self-aggregation and has been extensively studied since Held et al. (1993) (Bretherton et al. 2005; Muller and Held 2012; Wing and Emanuel 2014; Emanuel et al. 2014; Muller and Bony 2015; Holloway and Woolnough 2016; Yang 2018a,b, 2019). Previous studies have suggested that tropical cyclones (Nolan et al. 2007; Boos et al. 2016) and the Madden-Julian Oscillation (MJO) (Arnold and Randall 2015) are special forms of convective self-aggregation on the $f$ plane and the equatorial $\beta$ plane, respectively. Therefore, investigating the underlying physics of self-aggregation can give additional insights into such mysteries in tropical meteorology.

Previous studies have widely employed moist static energy (MSE) to diagnose convectively coupled circulations in the tropical atmosphere (Neelin and Held 1987; Kiranmayi and Maloney 2011; Andersen and Kuang 2012; Arnold et al. 2013; Pritchard and Yang 2016). For example, Bretherton et al. (2005) predicted the initial e-folding rate of self-aggregation based on a vertically integrated (VI) MSE budget. Following Andersen and Kuang (2012), Wing and Emanuel (2014) developed a budget equation for the spatial variance of the VI-MSE [their Eq. (9)]. This VI-MSE variance contains both the local MSE variance (LMSE variance) at a given altitude and the covariance of MSE anomalies between different altitudes. Therefore, we refer to this framework as the *global* MSE (GMSE) variance framework. The GMSE variance framework showed that the development of self-aggregation is associated with an increase in the VI-GMSE variance. Based on this framework, the authors further attributed self-aggregation to individual physical processes, including radiative feedbacks, surface-flux feedbacks, and atmospheric circulation. The VI-GMSE variance framework has then been widely used to study self-aggregation (Coppin and Bony 2015; Arnold and Randall 2015; Wing and Cronin 2016; Holloway and Woolnough 2016).

While the VI-GMSE variance framework has provided many insights into the physics of self-aggregation, the vertical dimension remains too physically important to be integrated over (e.g., Mapes 2016). For example, recent studies have shown that the boundary layer (BL) is particularly important in convective self-aggregation (Jeevanjee and Romps 2013; Muller and Bony 2015; Naumann et al. 2017; Yang 2018a,b, 2021). Muller and Bony (2015) found that the radiative cooling profiles, especially the low-level cooling in dry patches, affect self-aggregation. Yang (2018b) showed that the development of self-aggregation is associated with increases in the available potential energy (APE)—the energy reservoir for self-aggregation circulations. The author then proposed and showed that physical processes in the BL dominate the APE production and are, therefore, key to convective self-aggregation. This "bottom-up" development of self-aggregation cannot be understood by using the VI-GMSE variance framework, because it does not resolve the vertical dimension.

Here we propose a novel MSE variance framework that can resolve the vertical dimension to study the development of convective self-aggregation. This framework only contains the LMSE variance at individual vertical layers. This framework allows us to calculate the LMSE variance and its evolution at each individual vertical level (vertically resolved (VR) analysis), and to integrate over two arbitrary altitudes (VI analysis). Therefore, this LMSE variance framework complements the existing VI-GMSE variance framework and can help to test if the BL processes are key to the development of self-aggregation. Table 1 shows the overall structure of the paper. In section 2, we will introduce the model setup. In section 3, we will introduce the new LMSE variance framework with both VR and VI analysis. In section 4, we will use the LMSE variance framework to diagnose convective self-aggregation in a CRM control simulation. We will also compare our results with the APE analysis in Yang (2018b). In Section 5, we will compare the LMSE variance framework and the GMSE variance framework. Section 6 will summarize the main findings and discuss the implications.

## 2. Numerical model setup

We use the System for Atmospheric Modeling (SAM, version 6.10.8) to simulate convective self-aggregation. SAM is an anelastic CRM (Khairoutdinov & Randall, 2003) and has been widely used to simulate self-aggregation (Bretherton et al. 2005; Muller and Held 2012; Khairoutdinov and Emanuel 2013; Wing and Emanuel 2014; Bretherton and Khairoutdinov 2015; Wing et al. 2016; Muller and Romps 2018; Yang 2018a,b, 2019). The thermodynamic prognostic variables in SAM are liquid/ice water static energy ($h_L$), total nonprecipitating water mixing ratio ($q_n$), and total precipitating water mixing ratio ($q_p$). Here, $h_L = c_p T + gz - L_v(q_{cw} + q_r) - L_s(q_{ci} + q_s + q_g) = c_p T + gz - L_v(q_{liquid} + q_{ice}) - L_f q_{ice}$, where $c_p$ is the specific heat of air at constant pressure, $q_{liquid}$ (= cloud water + rain = $q_{cw} + q_r$) and $q_{ice}$ (= cloud ice + snow + graupel = $q_{ci} + q_s + q_g$) are the mixing ratios of all liquid and ice phase condensates, and $(L_v, L_s, L_f)$ are latent heat of evaporation, sublimation and fusion. $q_n \,(= q_{cw} + q_{ci})$ and $q_p \,(= q_r + q_s + q_g)$ are model outputs, and can be partitioned into the mixing ratios of different hydrometeors (e.g., cloud water) following SAM's temperature-based microphysics.

In this paper, we study a 2D (x-z) aggregation simulation over a horizontal periodic domain. The domain size is 2,048 km, and the horizontal resolution is 2 km. The model top is at 42.9 km. There are 80 vertical grid levels, and the first level is at 37.5 m. The vertical resolution gradually increases from 80 m near the surface to 400 m above 5 km and 1 km above 25 km. There is a 6-km sponge layer on the top of the model to damp out gravity wave reflection. The simulation runs for 150 days with a fixed uniform SST at 300K. The data is output hourly. The radiative transfer scheme is the same as that of the National Center for Atmospheric Research Community Atmosphere Model (NCAR CAM3; Collins et al., 2006). The incident solar shortwave radiation is fixed at 413.9 W/m² to represent the climatological solar insolation received by the tropics. Considering that convective self-aggregation is a slow process, we turn off diurnal variations in the model for simplification. The microphysics is the one-moment parameterization. SAM parameterizes subgrid-scale (SGS) process using a Smagorinsky-type parameterization for turbulent fluxes within the atmosphere and using the Monin–Obukhov similarity theory for the surface turbulent fluxes.

## 3. MSE theory

Following Bretherton et al. (2005), the frozen moist static energy (FMSE, $h$) is defined as

$$h = c_p T + gz + L_v q_v - L_f q_{ice}, \qquad (1)$$

where $q_v$ is the specific humidity of water vapor. FMSE is approximately conserved in moist adiabatic processes where pressure change is hydrostatic, with slight influence from liquid and ice condensates on the change of internal energy. We use this variable because it is exactly conserved in SAM and has been used to study convective self-aggregation in previous studies (Bretherton et al. 2005; Wing and Emanuel 2014; Carstens and Wing 2020). Hereafter, FMSE will be simplified as MSE.

Given that both the GMSE and LMSE frameworks can be derived from the MSE budget, we start our analysis from the two-dimension MSE budget equation. Its flux form is given by

$$\partial_t h = -\frac{1}{\rho_0}[\partial_x(\rho_0 u h) + \partial_z(\rho_0 w h)] + Q_{rad} + Q_{sgs}, \qquad (2)$$

where $\rho_0 = \rho_0(z)$ is the reference density, $(u, w)$ are resolved wind speeds, and $(Q_{rad}, Q_{sgs})$ are the MSE sources/sinks due to radiation and SGS processes, respectively (units: W kg$^{-1}$). Following the convention in SAM (Khairoutdinov and Randall 2003), the SGS MSE tendency is given by $Q_{sgs} = -\frac{1}{\rho_0}\partial_x F_H - \frac{1}{\rho_0}\partial_z F_V$, where $(F_H, F_V)$ are the horizontal and vertical turbulent fluxes (unites: W m$^{-2}$), and $F_V(z = 0)$ represents the surface turbulence fluxes of MSE $F_s$. Eq. (2) can be derived from the thermodynamic prognostic equations in SAM (A3-A5 in Khairoutdinov and Randall (2003)). Then, we take the horizontal average of

Eq. (2) and get

$$\overline{\partial_t h} = -\frac{1}{\rho_0}\overline{\partial_z(\rho_0 w h)} + \overline{Q_{rad}} + \overline{Q_{sgs}}. \quad (3)$$

The overbar denotes the horizontal average, and $-\frac{1}{\rho_0}\overline{\partial_x(\rho_0 u h)}$ vanishes in a periodic domain. $(x, z, t)$ are independent variables in SAM, and density only varies with height, so overbars can be moved into the differential symbols. Then we subtract Eq. (3) from Eq. (2), and get the MSE perturbation budget:

$$\partial_t h' = -\frac{1}{\rho_0}[\partial_x(\rho_0 u h) + \partial_z(\rho_0 w h)'] + Q'_{rad} + Q'_{sgs}, \quad (4)$$

where the superscripts $(')$ denote perturbations at resolved scales. Figure 1 further shows how diabatic/adiabatic processes change $h'$ in a two-layer model.

Because we are only interested in perturbations associated with large-scale convective aggregation (denoted as $\widetilde{A}'$, $A$ is a given variable), we apply both spatial (102-km) and temporal (5-day) running averages to filter out small-scale and high-frequency components. The results are robust with different smoothing windows (e.g. 22 km and 1 day; Appendix A). Then the budget equation for the large-scale MSE perturbations ($\widetilde{h}'$) is

$$\partial_t \widetilde{h}' = -\frac{1}{\rho_0}\left[\partial_x(\widetilde{\rho_0 u h}) + \partial_z(\widetilde{\rho_0 w h})'\right] + \widetilde{Q'_{rad}} + \widetilde{Q'_{sgs}}. \quad (5)$$

Hereafter, we denote the large-scale variable $\tilde{A}$ as $A$ for simplification.

3.1 The VI-GMSE variance framework

We first introduce a few useful definitions in the following analysis. The VI-MSE is defined as the total column MSE per unit area, which is

$$\hat{h} = \int_0^{z_t} \rho_0 h \, dz, \qquad (6)$$

where $\hat{A} = \int_0^{z_t} \rho_0 A \, dz$ denotes a density-weighted integral of variable $A$, and $z_t$ is the height of the model top. The VI-MSE is closely linked to the column water vapor due to the weak temperature gradient (WTG) in the tropics (Charney 1963; Sobel et al. 2001; Yang and Seidel 2020; Seidel and Yang 2020). Then the spatial variance of the VI-MSE (the VI-GMSE variance, $var_I$) is written as

$$var_I = \overline{(\hat{h}')^2} = \overline{\left(\int_0^{z_t} \rho_0 h \, dz - \overline{\int_0^{z_t} \rho_0 h \, dz}\right)^2} = \overline{\left(\int_0^{z_t} \rho_0 (h - \bar{h}) \, dz\right)^2} = \overline{\left(\int_0^{z_t} \rho_0 h' \, dz\right)^2}. \qquad (7)$$

We move the overbar into the integral sign because $\rho_0$ and $z$ are independent of $x$ in SAM. Following Wing and Emanuel (2014), the budget equation for the VI-GMSE variance is given by

$$\tfrac{1}{2} \partial_t (var_I) = \overline{\hat{h}' \cdot \left[-\tfrac{1}{\rho_0} \widehat{\partial_x (\rho_0 u h)}\right]} + \overline{\hat{h}' \cdot \widehat{Q'_{rad}}} + \overline{\hat{h}' \cdot \widehat{Q'_{sgs}}}, \qquad (8)$$

where $\widehat{Q_{sgs}} = -\widehat{\partial_x F_H} + F_V(z=0)$. If the horizontal component $\widehat{\partial_x F_H}$ is negligible, then we have

$$\widehat{Q_{sgs}} \approx F_V(z=0) \equiv F_s, \tag{9}$$

We shall see if this approximation is appropriate later (Figure A2). Note that the vertical flux convergence $-\frac{1}{\rho_0}\partial_z(\rho_0 wh)'$ vanishes when integrated from surface to the model top. Therefore, the VI-GMSE variance framework does not show how the vertical convergence contributes to convective self-aggregation. The VI-GMSE variance continually increases during the development of aggregation, and the key to the framework is to evaluate processes generating $var_I$.

Here, we provide physical intuition for Eq. (8) in a two-layer atmosphere (Figure 1). We assume that the thickness of each layer is $H$ and is horizontally uniform. Then the VI-GMSE variance can be written as

$$var_I = \overline{(\rho_0 h_1' H + \rho_0 h_2' H)^2} = \rho_0^2 H^2 \Big( \underbrace{\overline{h_1'^2} + \overline{h_2'^2}}_{\text{local variance}} + \underbrace{2\overline{h_1' h_2'}}_{\text{covariance}} \Big), \tag{10}$$

where the subscripts of $h'$ represent layer numbers. Eq. (10) contains two parts: the *local variance* within each layer (local in height), and the *covariance* between the layers. The local variance represents horizontal inhomogeneity, and larger values correspond to more aggregated states.

Similar 'local variance' and 'covariance' components also exist in terms on the right-hand side of Eq. (8). Take radiative production as an example:

$$\overline{\widehat{h'} \cdot \widehat{Q'_{rad}}} = \rho_0^2 H^2 \overline{\left( \underbrace{h'_1 Q'_{rad_1} + h'_2 Q'_{rad_2}}_{\text{local variance production}} + \underbrace{h'_1 Q'_{rad_2} + h'_2 Q'_{rad_1}}_{\text{covariance production}} \right)}, \qquad (11)$$

Equations (11) contains two parts: local variance production and covariance production. Local variance production is effective in modulating local MSE variance directly: cooling low-MSE air at the same level further reduces the MSE and increases the MSE variance at that level. We distinguish 'radiative cooling' from 'radiation'. Radiation describes radiative transfer processes in general and is a non-local process, but radiative cooling results from radiative flux divergence and *directly* changes local temperatures and can induce large-scale circulations. Covariance production is only responsible for the change of covariance. This is clearer by showing its budget equation (leaving behind $\rho_0^2 H^2$):

$$\partial_t \overline{(h'_1 h'_2)} = \overline{h'_1 [-\partial_x(uh)]_2} + \overline{h'_2 [-\partial_x(uh)]_1} + \overline{h'_1 Q'_{rad_1}} + \overline{h'_2 Q'_{rad_2}} + \overline{h'_1 Q'_{sgs_1}} + \overline{h'_2 Q'_{sgs_2}}.$$
(12)

Therefore, even though covariance is part of the VI-GMSE variance, covariance production does not directly contribute to changes of the local MSE variance or horizontal inhomogeneity at a given altitude.

We have illustrated the components of the VI-GMSE variance framework using a two-layer atmosphere. To resolve the vertical dimension, we will present two MSE analysis frameworks in this paper. We will first focus on deriving the LMSE variance framework (Section 3.2) and then apply it to a CRM simulation (Section 4). We will also derive a VR-GMSE variance framework and illustrate the differences between LMSE and GMSE variance frameworks (Section 5).

3.2 The LMSE variance framework

The VI-GMSE variance was computed by first vertically integrating MSE anomalies and then calculating the spatial variance of the VI-MSE. However, the LMSE variance framework does the opposite: we first compute the spatial variance of MSE at each altitude and then perform the vertical integration, which is

$$var_R(z_1, z_2) = \frac{1}{2} \int_{z_1}^{z_2} \overline{(\rho_0 h')^2}\, dz'. \qquad (13)$$

We simplify the large-scale variable $\widetilde{h'}$ to $h'$. This LMSE variance $var_R(z_1, z_2)$ represents the integrated *local* MSE variance between $z_1$ and $z_2$. Again, 'local' means at each vertical level. To get the VR budget for the LMSE variance, we multiply $\rho_0^2 h'$ on both sides of Eq. (5) and take a horizontal average on both sides, which yields

$$\frac{1}{2}\partial_t \overline{(\rho_0 h')^2} = \overline{\rho_0^2 h'\left[-\frac{1}{\rho_0}\partial_x(\rho_0 u h) - \frac{1}{\rho_0}\partial_z(\rho_0 w h)'\right]} + \overline{\rho_0^2 h' Q'_{rad}} + \overline{\rho_0^2 h' Q'_{sgs}}. \qquad (14)$$

Eq. (14) is the VR-LMSE variance budget, where variance production due to diabatic/adiabatic processes only changes LMSE variance at a given altitude. Note that the adiabatic production of the LMSE variance is done by dynamics and is represented explicitly in the LMSE variance framework, complementing the VI-GMSE variance framework. To simplify the equation, we let

$$-\frac{1}{\rho_0}\nabla\cdot(\rho_0\vec{v}h)' = \underbrace{-\frac{1}{\rho_0}\partial_x(\rho_0 uh)}_{\text{horizontal convergence}} \underbrace{-\frac{1}{\rho_0}\partial_z(\rho_0 wh)'}_{\text{vertical convergence}}. \qquad (15)$$

It is the sum of horizontal and vertical convergence of MSE flux anomalies, representing the total adiabatic production of $h'$ in Eq. (5). Therefore, the vertical redistribution of the MSE (e.g., convective transport) is now explicitly formulated in the LMSE variance framework.

Then we integrate (14) over $z_1$ and $z_2$:

$$\frac{1}{2}\int_{z_1}^{z_2}\partial_t\overline{(\rho_0 h')^2}dz = \int_{z_1}^{z_2}\overline{\rho_0^2 h'\left[-\frac{1}{\rho_0}\nabla\cdot(\rho_0\vec{v}h)'\right]}dz + \int_{z_1}^{z_2}\overline{\rho_0^2 h'Q'_{rad}}dz + \int_{z_1}^{z_2}\overline{\rho_0^2 h'Q'_{sgs}}dz.$$
(16)

Eq. (16) is the VI-LMSE variance budget. We further present a method to calculate the variance production from surface fluxes independently in Appendix B. Note that the integral does not change if the time derivative is moved outside the integral because altitude in SAM is independent of time. Be careful if model altitude changes with time. To illustrate more details in the early stage, we normalize Eq. (16) with the VI-LMSE variance $var_R(0, z_t)(= \frac{1}{2}\int_0^{z_t}\overline{(\rho_0 h')^2}dz)$, which yields

$$\underbrace{\frac{\frac{1}{2}\int_{z_1}^{z_2}\partial_t\overline{(\rho_0 h')^2}dz}{var_R(0,z_t)}}_{\text{growth rate}} = \underbrace{\frac{\int_{z_1}^{z_2}\overline{\rho_0^2 h'\left[-\frac{1}{\rho_0}\nabla\cdot(\rho_0\vec{v}h)\right]'}dz}{var_R(0,z_t)}}_{\text{adiabatic production}} + \underbrace{\frac{\int_{z_1}^{z_2}\overline{\rho_0^2 h'Q'_{rad}}dz + \int_{z_1}^{z_2}\overline{\rho_0^2 h'Q'_{sgs}}dz}{var_R(0,z_t)}}_{\text{diabatic production}}. \quad (17)$$

Eq. (17) is the normalized VI-LMSE variance budget. The left-hand-side term is the growth rate of the VI-LMSE variance, and the right-hand-side terms represent the adiabatic and diabatic productions of the LMSE variance. Diabatic processes including radiation and SGS processes produce/consume the LMSE variance by coupling with the MSE anomalies. For example, anomalous radiative cooling in drier (low-MSE) regions promote convective self-aggregation by further reducing MSE there and increasing the LMSE variance. Meanwhile, large-scale circulations and convection redistribute the MSE adiabatically, generating LMSE variance. Again, this adiabatic term includes both horizontal and vertical convergence of the MSE flux (e.g., convective transport). In the following analysis, we will compute the adiabatic term as the residual of Eq. (5), as done by Bretherton et al. (2005) and Wing and Emanuel (2014).

This LMSE variance framework is different from the VI-GMSE variance framework because it focuses on the increase of local MSE variance at individual vertical layers (Table 2). Therefore, the diabatic processes only change the MSE variance locally, and their remote effects are achieved by circulation and convection. Additionally, the budget equation of the LMSE variance can illustrate the vertical transport of the MSE and the vertical structure of convective self-aggregation. Therefore, the LMSE variance framework complements the VI-GMSE variance framework.

## 4. The LMSE variance diagnostic results

We use Eq. (14) and (17) to diagnose convective self-aggregation in a CRM simulation. We will show its evolution and vertical structure and will illustrate the importance of the BL.

Figure 2 plots the precipitable water (PW, mm) and the LMSE variance in the CRM simulation. Initially, three expending dry patches centered at $x =$ 400, 1300 and 2000 km start to form around day 15 (Figure 2a). During days 15-50, the system quickly evolves into a more aggregated state, with two dry patches merging into one heavily dry patch centered at $x =$1500 km. After day 50, the whole system reaches its statistical equilibrium, and PW variance oscillates around a reference value (Figure 2b). However, there is still some modulation within the system: one dry patch disappears when two moist patches merge into one patch on day 120. During the last 30 days, there is only one convective aggregate with a spatial scale of about 2000 km. This spatial scale is consistent with simulation results presented by Yang (2018a), who also provided a quantitative explanation for what sets the size of convective aggregates.

Figure 2b shows the VI-LMSE variance, which covaries with PW variance and describes the evolution of self-aggregation. The VI-LMSE variances increase together with the PW variance before day 50. After day 50, they all oscillate around their reference values. This is consistent with Figure 2a and Figure A3c, where the VI-LMSE variance also shares a similar tendency with the VI-GMSE variance. Their consistency confirms that the development of aggregation is associated with increases in the LMSE variances.

Figure 2b further shows that the LMSE variance is dominated by the variance in the BL (dashed line) and that the lower free troposphere (FT, dotted line) also contributes significantly to the LMSE variance, suggesting a bottom-heavy structure. Following Yang (2018a), we define the BL height as the altitude where $\partial_z RH$ first exceeds the threshold of $-0.1$ km$^{-1}$ over the level of the minimum gradient, which is about the lowest 2 km (Figure 3). Strong horizontal pressure gradient can no longer be sustained above that altitude. Figure 2c shows that much of the LMSE variance is within the lowest 2 km of the atmosphere, and that the LMSE variance is dominated by its moisture variance (Figure 2d). The bottom-heavy structure is likely because the water vapor mixing ratio exponentially decreases in altitude with a scale height of about 2 km. This result indicates that processes in the lower troposphere, especially in the BL, are important, because diabatic/adiabatic variance production can only become significant at layers where the MSE anomaly is large. We also notice a sharp increase of the LMSE variance at around 1 km. It is roughly the height of the cloud base. The relative humidity below is more horizontally uniform due to the efficient moistening by the underlying ocean and the strong vertical mixing within the sub-cloud layer. Therefore, large horizontal LMSE variations are absent below 1 km.

Figure 4 plots the normalized VI-LMSE variance budget integrated over different layers [see Eq. (17)] and further illustrates the importance of the BL. The green line represents the net growth rate of self-aggregation, which describes the temporal evolution of the VI-LMSE variance. A positive value suggests an overall tendency to aggregate. The yellow line measures the contribution of the adiabatic production of the VI-LMSE variance. A positive value indicates upgradient MSE transport, favoring aggregation. The dark blue line denotes variance production by radiative cooling and the purple one denotes that by SGS processes. Positive values suggest that the corresponding diabatic process increases the VI-LMSE variance and promotes aggregation. In Figure 4a, solid lines correspond to column integrals in Eq. (17) ($z_1 = 0, z_2 = z_t$), and dashed lines correspond to the BL integrals ($z_1 = 0, z_2 = z_{BL}$). Their differences measure the contribution of the FT, much of which is from the 2-4 km (Figure 4b). Note that the column curve and the BL curve for SGS coincide with each other. Initially, the VI-LMSE variance increases rapidly in both the BL and the lower FT. After dry patches form (around day 15), the BL starts to dominate the production of VI-LMSE variance, increasing the VI-LMSE variance. The lower FT only accounts for a small portion of column variance increase afterwards. This result is consistent with the LMSE variance plots in Figure 2b-c.

Figure 4 also supports the notion that dominant mechanisms in generating LMSE variance might be distinct in different stages of convective self-aggregation (Wing and Emanuel 2014). In the first 15 days, column adiabatic production, further enhanced by radiation, dominates the LMSE variance production for convective self-aggregation (solid lines), and both BL and lower FT processes make significant contributions to the organization of convection. After day 15, however, the production of LMSE variance becomes bottom-heavy and radiative processes replace adiabatic atmospheric circulation as the major source for variance production. During the whole simulation period, SGS processes are always responsible for consuming LMSE variance and inhibiting self-aggregation.

Figure 5 explicitly shows the diagnostic results of VR-LMSE variance budget [see Eq. (14)]. Figure 5a plots the *local* tendency of LMSE variance ($\partial_t \frac{1}{2}\overline{(\rho_0 h')^2}$), measuring how fast self-aggregation evolves at a given altitude. We observe a bottom-heavy structure after day 15: the LMSE tendency is primarily in the BL, the lowest 2 km of the atmosphere. There is positive variance tendency in the BL during day 15-50, suggesting the MSE anomalies are significantly intensified there. This is consistent with the bottom-heavy structure in Figure 4.

Figure 5b shows the production of the VR-LMSE variance by radiative cooling. Positive values suggest that radiative cooling anomalies are in phase with MSE anomalies and that radiative processes increase the LMSE variance, leading to self-aggregation. Again, the radiative LMSE variance production has a bottom-heavy structure: most production is within the lowest 2 km, which highlights the importance of BL. Diagnosis for the first 20 days is in white due to small magnitudes and will be discussed in Figure 6. During the simulation period, radiative processes continuously make a positive contribution to the development and the maintenance of convective self-aggregation. This result is consistent with our Figure 4 and the APE analysis and the mechanism-denial experiments in Yang (2018b).

Figure 5c measures the adiabatic production of the VR-LMSE variance by large-scale circulations and convection. Here we exclude the near-surface contribution due to its large magnitude and will discuss it in Figure 7. In Figure 5, the adiabatic production is the major process to balance the variance production by radiative cooling above the near-surface layer.

The adiabatic production has a similar bottom-heavy structure. Negative values represent downgradient MSE transport (e.g., MSE flux divergence in high-MSE regions), inhibiting self-aggregation. Generally, the adiabatic production above the near-surface layer is dominated by the BL and inhibits convective self-aggregation during days 15-50. There are also some positive values within the BL, which is likely due to the low-level circulation (Bretherton et al. 2005; Muller and Held 2012; Jeevanjee and Romps 2013; Coppin and Bony 2015).

Figure 5d shows the contribution from SGS processes. SGS production of the near-surface layer dominates the entire SGS production throughout the column. Therefore, the near-surface layer is excluded and will be discussed in Figure 6. Even though the SGS contribution is dominated by the near-surface layer (the lowest atmosphere level), SGS contribution remains significant below 200 m.

Figure 6 illustrates the vertical structures of variance tendency, radiative production, adiabatic production and SGS production in the first 20 days. It shows that the LMSE variance increases both in the lower FT and the BL during first 15 days, suggesting that both the lower FT and the BL are important for the development of convective self-aggregation. A downward propagation of maximum variance tendency occurs from the lower FT to the BL from day 15. We can also find a similar downward propagation in radiative production (Figure 6b) and adiabatic production (Figure 6c). These may suggest that there is an interaction between the BL and the FT.

Figure 7a-d gives more details on the roles of MSE flux convergence and SGS processes at the near-surface layer (the lowest model level). Figure 7a shows that there is a competing mechanism at the lowest level: MSE flux convergence is in phase with the near-surface MSE anomalies (Figure 7b-c) and continuously generates LMSE variance, while SGS anomalies (Figure 7d) have the opposite phase to the MSE anomalies and consume LMSE variance. Note that the magnitude of $Q'_{sgs}$ is two orders of magnitude larger than that of $Q'_{rad}$ at the near-surface layer. In radiative-convective equilibrium, VI SGS tendencies approximately equal surface fluxes (Figure A2), which balance the total radiative cooling of the whole column. $Q'_{sgs}$ has extreme large values at the near-surface layer due to surface fluxes, and the radiative cooling rate cannot change much locally, so adiabatic MSE sinks/sources must be large enough to balance the considerable SGS tendencies at the near-surface layer. Note that this positive adiabatic production at the near-surface layer is so large that it dominates over the negative adiabatic contribution above (Figure 5c).

## 5. Comparison between the LMSE and the GMSE variance frameworks

Here we compare the LMSE and GMSE frameworks. Section 3 has developed both the VI- and VR-LMSE variance analyses. To conduct a comprehensive comparison, we first derive a VR-GMSE variance budget and then compare the two frameworks from both the VI and VR perspectives (Tables 1 & 2).

The VR-GMSE variance equation is derived by first multiplying $\rho_0 \widehat{h'}$ and then taking the horizontal average on both sides of Eq. (5), which is

$$\overline{\widehat{h'} \cdot \partial_t \rho_0 h'} = \overline{\widehat{h'} \cdot [-\partial_x(\rho_0 uh) - \partial_z(\rho_0 wh)']} + \overline{\widehat{h'} \cdot \rho_0 Q'_{rad}} + \overline{\widehat{h'} \cdot \rho_0 Q'_{sgs}}. \quad (18)$$

We can integrate Eq. (18) between any altitudes and quantify variance production within that layer. If we integrate Eq. (18) from the surface to the model top, *the VR-GMSE budget reduces to the VI -GMSE budget equation* [Eq. (8)].

Figure 8 shows the VI- and VR-GMSE variance diagnostic results. The VI-GMSE variance analysis reduces to the conventional MSE analysis (e.g., Figure 5 in Wing and Emanuel 2014). Figure 8a shows the normalized VI-GMSE budget, calculated using Eq. (8) divided by the VI-GMSE variance. The result shows that radiation dominates the VI-GMSE variance production during the 150-day simulation. Surface fluxes and horizontal convergence of MSE flux help to promote the development of self-aggregation in the first 15 days. After day 15, surface fluxes become the major processes that consume GMSE variance and inhibit self-aggregation. Figure 8a also shows that both variance tendency and variance productions are dominated by the BL processes after day 15, while the FT also makes an important contribution in the first 15 days. Such bottom-heavy structure after day 15 is further confirmed by the VR-GMSE variance diagnosis shown in Figure 8b-c. Radiation generally increases the GMSE variance at individual layers, while the adiabatic processes consume the GMSE variance in most layers.

There are high similarities between the LMSE and GMSE variance frameworks. For example, the VI budgets in the two frameworks agree that both the BL and the lower FT make important contributions in the first 15 days. After day 15, the VI- and the VR- budgets in the two frameworks consistently highlight the key role of BL processes in the development of self-aggregation. The signs and structures of variance tendency and variance productions in the two frameworks are also generally identical after day 15. Such consistency confirms the robustness of the bottom-heavy results.

Meanwhile, there are still notable differences between the two frameworks in the first 15 days. For example, the VI-LMSE variance production is dominated by adiabatic production, while the VI-GMSE variance production is dominated by radiation. Moreover, the BL adiabatic productions have opposite signs between the two VI diagnoses in the first 15 days. The SGS production also has opposite signs in the two VI diagnoses: slightly negative in the VI-LMSE variance diagnosis, and strongly positive in the VI-GMSE variance diagnosis. Such differences may result from the inclusion of the covariance term in the GMSE framework (Table 2). We have summarized the differences in the mathematical formulation for both VI and VR budgets between the two frameworks in Table 2.

There are caveats in both diagnostic frameworks. For example, previous research showed that low-level circulations promote the development of convective self-aggregation (Bretherton et al. 2005, Jeevanjee and Romps 2013, Coppin and Bony 2015). However, the BL adiabatic production in the VI-GMSE variance diagnosis is strongly negative in the first 15 days, inhibiting self-aggregation. In addition, Bretherton et al. (2005) showed that self-aggregation did not occur with horizontally homogenized surface fluxes. However, the SGS production in the VI-LMSE variance diagnosis is a slightly negative term in the first 15 days.

We then perform a mechanism-denial experiment to better understand the role of SGS processes, including surface fluxes. In this experiment, we homogenize surface fluxes horizontally at each time step (Figure A4). The atmosphere quickly self-aggregates into moist and dry patches in the first 15 days (Figure A4b), and the horizontal moisture contrast reaches a similar magnitude to that in the control simulation (Figure A4a). This implies that surface-flux feedbacks have limited impacts on the development of self-aggregation in our simulations. This result agrees with the LMSE variance diagnosis, which shows weakly negative SGS production in the first 15 days. This mechanism-denial experiment seems to contradict with the GMSE variance diagnosis, which shows strongly positive SGS production in the first 15 days.

Figure A3 compares the difference between the GMSE framework and our LMSE framework using a CRM simulation. The diagonal components represent *local* variance and *local* variance production. The off-diagonal components represent the covariance and covariance production. Therefore, the VI-GMSE variance budget [Eq. (8)] includes every point in the matrix while the VI-LMSE variance budget [Eq. (16)] only cares about points on the diagonal. If we focus on the VR budgets and take radiative production at 4 km as an example (Figure A3b), the variance production in the VR-GMSE variance budget [Eq. (18)] is the sum of every point on the horizontal line, containing both the local variance and the covariance, while the variance production in the VR-LMSE variance budget [Eq. (14)] is just the red point on the diagnal of the matrix.

## 6. Conclusion and Discussion

Previous studies have extensively used a vertically integrated (VI) moist static energy (MSE) framework to study the development of convective self-aggregation. The framework technically diagnoses aggregation as the increase of the spatial variance of the VI-MSE. In this paper, we first decompose this framework with a two-layer model. We find that its MSE variance contains both the spatial variance of the local MSE anomalies at individual layers (LMSE variance) and the covariance of the MSE anomalies between different layers [Eq. (10)]. Therefore, we refer to this VI framework as the global MSE (GMSE) variance framework. We then show that the LMSE variance is only produced by the local variance production defined in Eq. (11), and that the covariance is only changed by covariance production, which does not contribute to changes in the LMSE variance [Eq. (12)].

To illustrate the vertical structure of convective self-aggregation, here we present a vertically resolved (VR) MSE variance framework that focuses on changes of the LMSE variance [Eq. (14)]. We first show that the LMSE variance has a bottom-heavy structure, which is likely due to the exponential decrease of water vapor with altitude (Figure 2). We further show that both diabatic and adiabatic productions of LMSE variance share a similar bottom-heavy structure, leading to the primary increase of LMSE variance in the BL. Besides, the lower FT also contributes to the increase of the LMSE variance in the first 15 days. These results are consistent with the previous APE analysis and mechanism-denial experiments and support the hypothesis that physical processes in the lower troposphere, especially the BL, are key to self-aggregation (Yang 2018b; Muller and Bony 2015; Naumann et al. 2017).

The bottom-heavy structure of the LMSE variance suggests that lower troposphere is important to the development of convective organizations. This is not only true for self-aggregation in the CRMs, but is also consistent with theoretical reasoning and observations. For example, previous studies showed the onset of deep convection is closely related to the lower troposphere moisture variability (Tompkins 2001; Parsons et al. 2000; Brown and Zhang 1997; Holloway and Neelin 2009) and that low-to-mid level moistening can induce moisture–stratiform instability for convectively coupled circulations (Mapes 2000; Kuang 2008). Meanwhile, Parker et al. (2016) showed that the onset of monsoon systems is associated with the moistening of the lower troposphere. These results on different spatial scales all illustrate the importance of the lower troposphere MSE variabilities. Therefore, the LMSE variance framework can be used to diagnose the evolution of such real-world convection and illustrate the vertical structure. This may in turn give further physical intuition to convective parameterization in the general circulation models.

We find that adiabatic processes favor self-aggregation in the first 50 days (yellow lines in Figure 4a). It seems consistent with previous studies, which proposed that the upgradient MSE transport by the low-level circulation favors self-aggregation (Bretherton et al. 2005; Muller and Bony 2015). However, here we demonstrate that the adiabatic MSE production is mainly in the near-surface layer (Figure 7a). This is because near-surface MSE flux convergence needs to be large enough to balance the considerable contribution from SGS processes, which cannot be achieved by local radiative cooling. This is a unique result of the LMSE variance framework, which explicitly resolves the vertical dimension.

The diagnostic results of the GMSE and the LMSE variance frameworks can be different. A reason is that the GMSE variance production is dominated by covariance production — the off-diagonal terms in Figure A3, but the LMSE variance production only includes the diagonal terms. Suppose we have enhanced surface fluxes moisten a dry surface layer below a moist column. The LMSE framework will show a reduction in LMSE variance because the enhanced surface fluxes moisten the dry layer. The indirect influence of surface fluxes is through convection and large-scale circulations, whose contribution is explicitly calculated as adiabatic LMSE variance production. In other words, the convective fluxes in the LMSE variance framework communicates the influence of surface enthalpy fluxes upwards. In contrast, the GMSE variance framework will diagnose the moistening on the dry layer as positive contribution. This contrast may result from the inclusion of covariance terms in the GMSE framework (Table 2). This GMSE framework does not explicitly calculate the vertical transport of the MSE flux and instead includes a covariance term (e.g., the off-diagonal components in Figure A3b). The covariance terms did not receive much attention in previous research, and its physical meaning is still unclear. Therefore, future work is needed to understand the covariance terms and to reconcile the difference between LMSE and GMSE variance frameworks.

Recall that the VI-LMSE variance diagnosis shows the SGS contribution is negative in the first 15 days (Figure 4a). This result seems to be contrary with previous mechanism-denial experiments showing that convection does not self-aggregate in simulations without interactive surface energy fluxes (e.g., Bretherton et al., 2005; Holloway and Woolnough 2016). To reconcile our results with previous studies, we performed a mechanism-denial experiment, in which we horizontally homogenize surface energy fluxes at each time step (Figure A4b). In our mechanism-denial experiment, the atmosphere quickly self-aggregates into a moist patch and a dry patch. The horizontal moisture contrast reaches a similar or even higher magnitude to that in our control simulation (Figure A4a). This simulation result agrees with the modest, negative SGS contribution from the VR-MSE diagnosis. However, one may ask, what makes our mechanism-denial experiment different from previous ones (e.g., Bretherton et al. (2005))? We suspect that most previous studies used small-domain simulations, in which the degree of convective aggregation is sensitive to model parameters and setups. In future studies, we plan to explore this hypothesis by performing simulations with different domain sizes.

Even though it can be insightful to illustrate the vertical structure, the LMSE variance framework only diagnoses convective organizations from a thermodynamic perspective. It will be helpful to compare this framework with other vertically resolved diagnostic methods based on energy conservation or dynamics (e.g., the APE framework). Additionally, we calculate the adiabatic term as a residual in this paper. It would be desirable to compute this term directly with frequent model output or online diagnosis and to illustrate the vertical structures of the horizontal and vertical adiabatic production (Figure 3 in Wolding et al. (2016)).

We plan to apply the LMSE variance framework to study other convectively coupled circulations, such as the MJO. The MJO has a rich vertical structure in winds, temperature, moisture, and clouds. However, previous studies have primarily used the VI-GMSE variance framework to investigate the evolution of the MJO, which ignores the vertical structure (Andersen and Kuang 2012; Arnold and Randall 2015; Pritchard and Yang 2016; Kiranmayi and Maloney 2011; Maloney 2009; Sobel et al. 2014). Wolding et al. (2016) assumed a weak horizontal temperature gradient and developed a vertically resolved analysis method for the MJO. That framework may work well in the free troposphere but introduces uncertainties in the boundary layer, where substantial horizontal temperature gradient can be sustained. Therefore, our LMSE variance analysis will complement the previous studies and help understand the MJO evolution in time and altitude.

**Acknowledgment:**

This work was supported by the Laboratory Directed Research and Development (LDRD) funding from Berkeley Lab, provided by the Director, Office of Science, of the U.S. Department of Energy under Contract DE-AC02-05CH11231, by the U.S. Department of Energy, Office of Science, Office of Biological and Environmental Research, Climate and Environmental Sciences Division, Regional & Global Climate Modeling Program, under Award DE-AC02-05CH11231, and by the Packard Fellowship for Science and Engineering (to D.Y.). The authors thank Dr. P. O'Gorman and anonymous reviewers for their constructive comments and helpful suggestions.

# APPENDIX A

**Sensitivity to choices of smoothing window widths**

Here we apply various temporal (1, 3 and 5 days) and spatial (22, 50, 82 and 102 km) smoothing to test if the VI-LMSE variance diagnosis is sensitive to the choices of window widths. The normalized plots are also smoothed with a 5-day running average to filter out the high-frequency signals. The results are shown in Figure A1. The blue box labels the window widths we use in the paper. Figure A1 confirms that the diagnostic results are robust under smaller spatial windows.

APPENDIX B

**Diabatic production of the LMSE variance due to surface fluxes**

Surface-flux contribution to the LMSE variance production is included as a component of the SGS production (see Eq. (9) and Figure A2). Here, we isolate the surface-flux contribution and discuss how to calculate this component. In Eq. (16), the SGS production is given by $\int_{z_1}^{z_2} \overline{\rho_0^2 h' Q'_{sgs}} dz$. Similarly, the variance production from surface fluxes is given by

$$\text{variance production} = \int_0^{z_2} \overline{\rho_0^2 h' Q'_{sfc}} dz, \qquad (A1)$$

where $Q'_{sfc} = -\frac{1}{\rho_0}\partial_z F'_s$. $F_s$ is the surface enthalpy fluxes in the unit of $\text{W m}^{-2}$ and is directly calculated and output by the model using the bulk formula. Integrating Eq. (A1) by parts yields

$$\int_0^{z_2} \overline{\rho_0 h'(-\partial_z F'_s)} dz = -\int_0^{z_2} \overline{\partial_z (\rho_0 h' F'_s)} dz + \int_0^{z_2} \overline{F'_s \partial_z (\rho_0 h')} dz. \qquad (A2)$$

Because $F_s$ is non-zero only at the surface, the second term on the right-hand side vanishes. Then we get

$$\text{variance production} = \overline{(\rho_0 h' F_s')}|_{surface} = \overline{\rho_{0s} h_s' F_s'}. \quad (A3)$$

We can use values at the lowest atmosphere level to calculate the numerical values of Eq. (A3).

Another method to calculate the variance production is:

$$\text{variance production} = \int_0^{Z_2} \overline{\rho_0 h'(-\partial_z F_s')} dz = \overline{\rho_{0s} h_s' \frac{F_s'}{\Delta z}} \Delta z = \overline{\rho_{0s} h_s' F_s'}. \quad (A4)$$

Eq. (A3) and (A4) are consistent.

**REFERENCES**

Andersen, J. A., and Z. Kuang, 2012: Moist static energy budget of MJO-like disturbances in the atmosphere of a zonally symmetric aquaplanet. *J. Clim.*, **25**, 2782–2804, https://doi.org/10.1175/JCLI-D-11-00168.1.

Arnold, N. P., and D. A. Randall, 2015: Global-scale convective aggregation: Implications for the Madden-Julian Oscillation. *J. Adv. Model. Earth Syst.*, **7**, 1499–1518, https://doi.org/10.1002/2015MS000498.

——, Z. Kuang, and E. Tziperman, 2013: Enhanced MJO-like variability at high SST. *J. Clim.*, **26**, 988–1001, https://doi.org/10.1175/JCLI-D-12-00272.1.

Boos, W. R., A. Fedorov, and L. Muir, 2016: Convective self-aggregation and tropical cyclogenesis under the hypohydrostatic rescaling. *J. Atmos. Sci.*, **73**, 525–544, https://doi.org/10.1175/JAS-D-15-0049.1.

Bretherton, C. S., and M. F. Khairoutdinov, 2015: Convective self-aggregation feedbacks in near-global cloud-resolving simulations of an aquaplanet. *J. Adv. Model. Earth Syst.*, **7**, 1765–1787, https://doi.org/10.1002/2015MS000499.

——, P. N. Blossey, and M. Khairoutdinov, 2005: An energy-balance analysis of deep convective self-aggregation above uniform SST. *J. Atmos. Sci.*, **62**, 4273–4292, https://doi.org/10.1175/JAS3614.1.

Brown, R. G., and C. Zhang, 1997: Variability of midtropospheric moisture and its effect on cloud-top height distribution during TOGA COARE. *J. Atmos. Sci.*, **54**, 2760–2774, https://doi.org/10.1175/1520-0469(1997)054<2760:VOMMAI>2.0.CO;2.

Carstens, J. D., and A. A. Wing, 2020: Tropical cyclogenesis from self-aggregated convection in numerical simulations of rotating radiative-convective equilibrium. *J. Adv. Model. Earth Syst.*, **12**, https://doi.org/10.1029/2019MS002020.

Charney, J. G., 1963: A note on large-scale motions in the Tropics. *J. Atmos. Sci.*, **20**, 607–609, https://doi.org/10.1175/1520-0469(1963)020<0607:anolsm>2.0.co;2.

Collins, W. D., and Coauthors, 2006: The formulation and atmospheric simulation of the Community Atmosphere Model version 3 (CAM3). *J. Clim.*, **19**, 2144–2161, https://doi.org/10.1175/JCLI3760.1.

Coppin, D., and S. Bony, 2015: Physical mechanisms controlling the initiation of convective self-aggregation in a General Circulation Model. *J. Adv. Model. Earth Syst.*, **7**, 2060–2078, https://doi.org/10.1002/2015MS000571.

Emanuel, K., A. A. Wing, and E. M. Vincent, 2014: Radiative-convective instability. *J. Adv. Model. Earth Syst.*, **6**, 75–90, https://doi.org/10.1002/2013MS000270.

Holloway, C. E., and D. J. Neelin, 2009: Moisture vertical structure, column water vapor, and tropical deep convection. *J. Atmos. Sci.*, **66**, 1665–1683, https://doi.org/10.1175/2008JAS2806.1.

Holloway, C. E., and S. J. Woolnough, 2016: The sensitivity of convective aggregation to diabatic processes in idealized radiative-convective equilibrium simulations. *J. Adv. Model. Earth Syst.*, **8**, 166–195, https://doi.org/10.1002/2015MS000511.

Jeevanjee, N., and D. M. Romps, 2013: Convective self-aggregation, cold pools, and domain size. *Geophys. Res. Lett.*, **40**, 994–998, https://doi.org/10.1002/grl.50204.

Khairoutdinov, M., and K. Emanuel, 2013: Rotating radiative-convective equilibrium simulated by a cloud-resolving model. *J. Adv. Model. Earth Syst.*, **5**, 816–825,

https://doi.org/10.1002/2013ms000253.

Khairoutdinov, M. F., and D. A. Randall, 2003: Cloud resolving modeling of the ARM summer 1997 IOP: Model formulation, results, uncertainties, and sensitivities. *J. Atmos. Sci.*, **60**, 607–625, https://doi.org/10.1175/1520-0469(2003)060<0607:CRMOTA>2.0.CO;2.

Kiranmayi, L., and E. D. Maloney, 2011: Intraseasonal moist static energy budget in reanalysis data. *J. Geophys. Res. Atmos.*, **116**, 711–729, https://doi.org/10.1029/2011JD016031.

Kuang, Z., 2008: A moisture-stratiform instability for convectively coupled waves. *J. Atmos. Sci.*, **65**, 834–854, https://doi.org/10.1175/2007JAS2444.1.

Maloney, E. D., 2009: The moist static energy budget of a composite tropical intraseasonal oscillation in a climate model. *J. Clim.*, **22**, 711–729, https://doi.org/10.1175/2008JCLI2542.1.

Mapes, B. E., 2000: Convective inhibition, subgrid-scale triggering energy, and stratiform instability in a toy tropical wave model. *J. Atmos. Sci.*, **57**, 1515–1535, https://doi.org/10.1175/1520-0469(2000)057<1515:CISSTE>2.0.CO;2.

Mapes, B. E., 2016: Gregarious convection and radiative feedbacks in idealized worlds. *J. Adv. Model. Earth Syst.*, **8**, 1029–1033, https://doi.org/10.1002/2016MS000651.

Muller, C., and S. Bony, 2015: What favors convective aggregation and why? *Geophys. Res. Lett.*, **42**, 5626–5634, https://doi.org/10.1002/2015GL064260.

Muller, C. J., and I. M. Held, 2012: Detailed investigation of the self-aggregation of convection in cloud-resolving simulations. *J. Atmos. Sci.*, **69**, 2551–2565,

    https://doi.org/10.1175/JAS-D-11-0257.1.

——, and D. M. Romps, 2018: Acceleration of tropical cyclogenesis by self-aggregation feedbacks. *Proc. Natl. Acad. Sci. U. S. A.*, **115**, 2930–2935, https://doi.org/10.1073/pnas.1719967115.

Naumann, A. K., B. Stevens, C. Hohenegger, and J. P. Mellado, 2017: A conceptual model of a shallow circulation induced by prescribed low-level radiative cooling. *J. Atmos. Sci.*, **74**, 3129–3144, https://doi.org/10.1175/JAS-D-17-0030.1.

Neelin, J. D., and I. M. Held, 1987: Modeling tropical convergence based on the moist static energy budget. *Mon. Weather Rev.*, **115**, 3–12, https://doi.org/10.1175/1520-0493(1987)115<0003:MTCBOT>2.0.CO;2.

Nolan, D. S., E. D. Rappin, and K. A. Emanuel, 2007: Tropical cyclogenesis sensitivity to environmental parameters in radiative-convective equilibrium. *Q. J. R. Meteorol. Soc.*, **133**, 2085–2107, https://doi.org/10.1002/qj.170.

Parker, D. J., P. Willetts, C. Birch, A. G. Turner, J. H. Marsham, C. M. Taylor, S. Kolusu, and G. M. Martin, 2016: The interaction of moist convection and mid-level dry air in the advance of the onset of the Indian monsoon. *Q. J. R. Meteorol. Soc.*, **142**, 2256–2272, https://doi.org/10.1002/qj.2815.

Parsons, D. B., J. L. Redelsperger, and K. Yoneyama, 2000: The evolution of the tropical western Pacific atmosphere-ocean system following the arrival of a dry intrusion. *Q. J. R. Meteorol. Soc.*, **126**, 517–548, https://doi.org/10.1002/qj.49712656307.

Pritchard, M. S., and D. Yang, 2016: Response of the superparameterized Madden-Julian oscillation to extreme climate and basic-state variation challenges a moisture mode view.

*J. Clim.*, **29**, 4995–5008, https://doi.org/10.1175/JCLI-D-15-0790.1.

Seidel, S. D., and D. Yang, 2020: The lightness of water vapor helps to stabilize tropical climate. *Sci. Adv.*, **6**, eaba1951, https://doi.org/10.1126/sciadv.aba1951.

Sobel, A., S. Wang, and D. Kim, 2014: Moist static energy budget of the MJO during DYNAMO. *J. Atmos. Sci.*, **71**, 4276–4291, https://doi.org/10.1175/JAS-D-14-0052.1.

Sobel, A. H., J. Nilsson, and L. M. Polvani, 2001: The weak temperature gradient approximation and balanced tropical moisture waves. *J. Atmos. Sci.*, **58**, 3650–3665, https://doi.org/10.1175/1520-0469(2001)058<3650:TWTGAA>2.0.CO;2.

Tompkins, A. M., 2001: Organization of tropical convection in low vertical wind shears: The role of water vapor. *J. Atmos. Sci.*, **58**, 529–545, https://doi.org/10.1175/1520-0469(2001)058<0529:OOTCIL>2.0.CO;2.

Wing, A. A., and K. A. Emanuel, 2014: Physical mechanisms controlling self-aggregation of convection in idealized numerical modeling simulations. *J. Adv. Model. Earth Syst.*, **6**, 59–74, https://doi.org/10.1002/2013MS000269.

——, and T. W. Cronin, 2016: Self-aggregation of convection in long channel geometry. *Q. J. R. Meteorol. Soc.*, **142**, 1–15, https://doi.org/10.1002/qj.2628.

——, S. J. Camargo, and A. H. Sobel, 2016: Role of radiative-convective feedbacks in spontaneous tropical cyclogenesis in idealized numerical simulations. *J. Atmos. Sci.*, **73**, 2633–2642, https://doi.org/10.1175/JAS-D-15-0380.1.

Wolding, B. O., E. D. Maloney, and M. Branson, 2016: Vertically resolved weak temperature gradient analysis of the Madden-Julian Oscillation in SP-CESM. *J. Adv. Model. Earth Syst.*, **8**, 1586–1619, https://doi.org/10.1002/2016MS000724.

Yang, D., 2018a: Boundary layer height and buoyancy determine the horizontal scale of convective self-aggregation. *J. Atmos. Sci.*, **75**, 469–478, https://doi.org/10.1175/JAS-D-17-0150.1.

——, 2018b: Boundary layer diabatic processes, the virtual effect, and convective self-aggregation. *J. Adv. Model. Earth Syst.*, **10**, 2163–2176, https://doi.org/10.1029/2017MS001261.

——, 2019: Convective heating leads to self-aggregation by generating available potential energy. *Geophys. Res. Lett.*, **46**, 10687–10696, https://doi.org/10.1029/2019GL083805.

——, 2021: A shallow-water model for convective self-aggregation. *J. Atmos. Sci.*, **78**, 571–582, https://doi.org/10.1175/JAS-D-20-0031.1.

——, and S. D. Seidel, 2020: The incredible lightness of water vapor. *J. Clim.*, **33**, 2841–2851, https://doi.org/10.1175/jcli-d-19-0260.1.

|     | **LMSE** | **GMSE** |
| --- | --- | --- |
| **VI** | *This paper* (Sections 3.2-4) | Andersen and Kuang (2012) Wing and Emanuel (2014) |
| **VR** | *This paper* (Sections 3.2-4) | *This paper* (Section 5) |

**Table 1.** The overall structure of the paper. LMSE represents the local MSE variance framework, and GMSE represents the global MSE variance framework. VI represents vertically integrated analysis, and VR represents vertically resolved analysis.

| | **LMSE variance** | **GMSE variance** |
|---|---|---|
| **Integrated MSE variance** | $\frac{1}{2}\underbrace{\left(\overline{h_1'^2} + \overline{h_2'^2}\right)}_{\text{local variance}}$ | $\underbrace{\overline{h_1'^2} + \overline{h_2'^2}}_{\text{local variance}} + \underbrace{\mathbf{2\overline{h_1'h_2'}}}_{\text{covariance}}$ |
| **MSE variance at layer 1** | $\frac{1}{2}\overline{h_1'^2}$ | $\overline{h_1'^2} + \mathbf{\overline{h_1'h_2'}}$ |
| **Integrated adiabatic production** | $\underbrace{\overline{h_1'\left[-\frac{1}{\rho_0}\nabla\cdot(\rho_0\vec{v}h)'\right]_1} + \overline{h_2'\left[-\frac{1}{\rho_0}\nabla\cdot(\rho_0\vec{v}h)'\right]_2}}_{\text{local variance production from }\textbf{horizontal and vertical convergence}}$ | $\underbrace{\overline{h_1'[-\partial_x(uh)]_1} + \overline{h_2'[-\partial_x(uh)]_2}}_{\substack{\text{local variance production from}\\ \textbf{horizontal}\text{ convergence}}}$ $+ \underbrace{\mathbf{\overline{h_2'[-\partial_x(uh)]_1} + \overline{h_1'[-\partial_x(uh)]_2}}}_{\substack{\textbf{covariance production from}\\ \textbf{horizontal convergence}}}$ |
| **Adiabatic production at layer 1** | $\overline{h_1'\left[-\frac{1}{\rho_0}\nabla\cdot(\rho_0\vec{v}h)'\right]_1}$ | $\overline{h_1'\left[-\frac{1}{\rho_0}\nabla\cdot(\rho_0\vec{v}h)'\right]_1} + \mathbf{\overline{h_2'\left[-\frac{1}{\rho_0}\nabla\cdot(\rho_0\vec{v}h)'\right]_1}}$ |
| **Integrated diabatic production** | $\overline{h_1'Q_1'} + \overline{h_2'Q_2'}$ | $\overline{h_1'Q_1'} + \overline{h_2'Q_2'} + \mathbf{\overline{h_1'Q_2'} + \overline{h_2'Q_1'}}$ |
| **Diabatic production at layer 1** | $\overline{h_1'Q_1'}$ | $\overline{h_1'Q_1'} + \mathbf{\overline{h_2'Q_1'}}$ |

**Table 2.** Formulation in the LMSE and the GMSE variance frameworks in a two-layer model. Differences are marked in bold fonts.

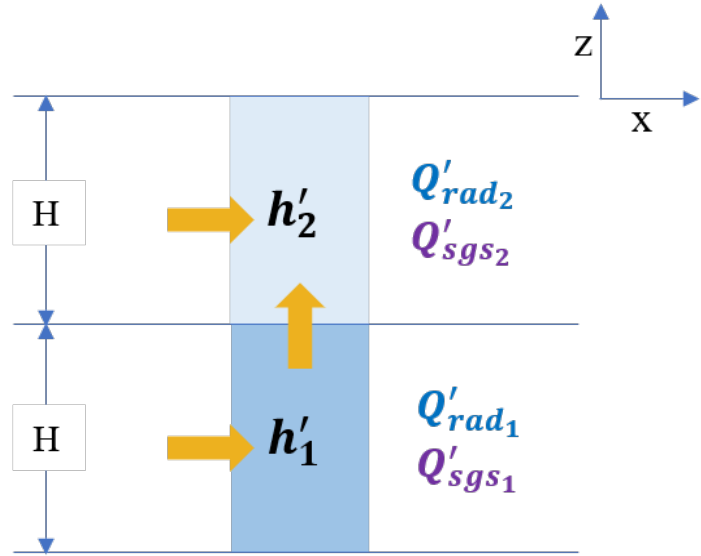

**Figure 1.** The two-layer schematic of convective self-aggregation. The orange arrows denote horizontal and vertical convergence of MSE fluxes. $\left(h', Q'_{rad}, Q'_{sgs}\right)$ denote perturbations of the MSE and MSE tendency from radiation and SGS processes, respectively. The units of $Q$ are all $W\ kg^{-1}$. The subscripts 1 and 2 represent layer numbers. The height for each layer H is constant.

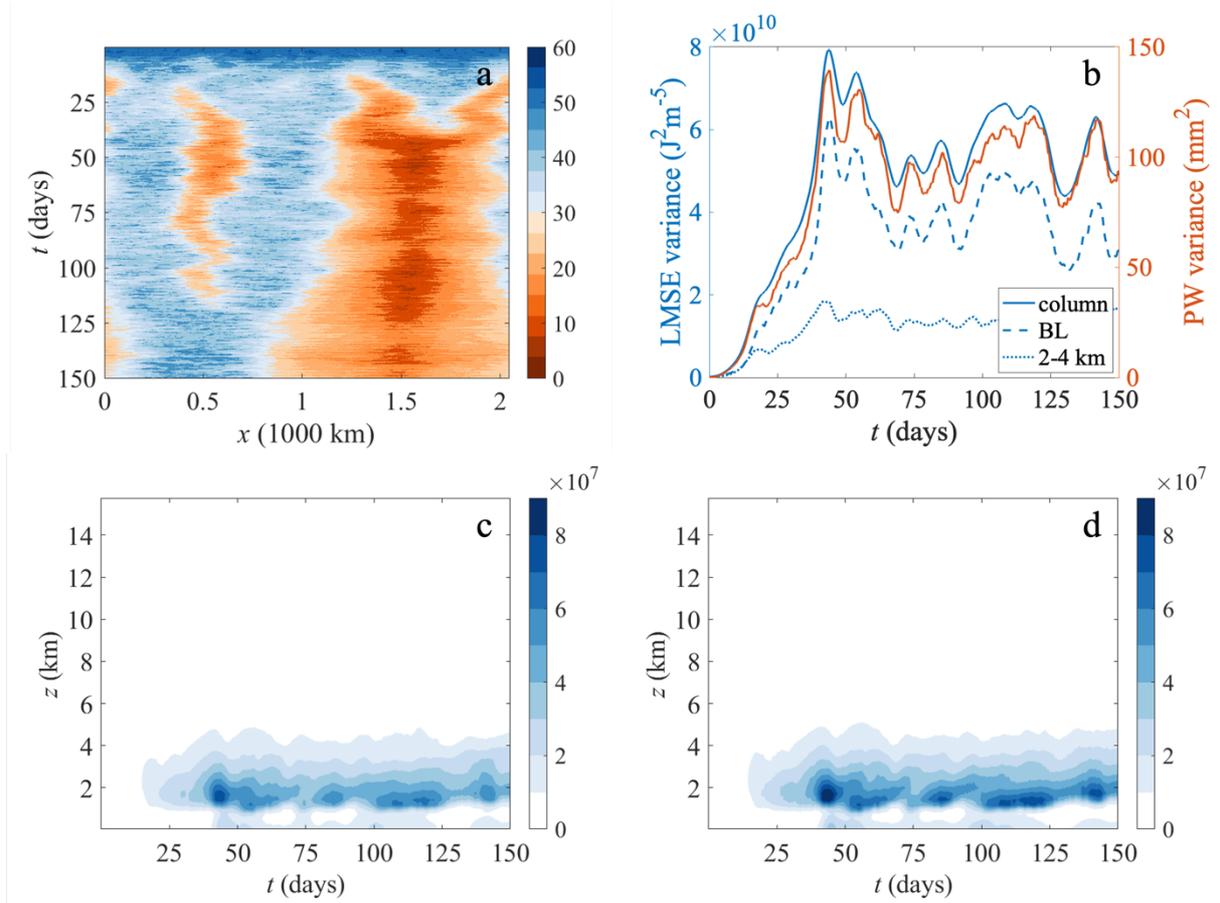

**Figure 2.** Evolution of precipitable water (PW, mm) and the LMSE variance. (a) Hovmöller diagram of the PW in the CRM simulation (blue: moist, orange: dry). (b) The evolution of the LMSE variance ($var_R(z_1, z_2)$, blue lines) and the PW variance (orange line). The solid blue line denotes the VI-LMSE variance over the column ($var_R(0, z_t)$), the dashed one denotes the variance within the BL (0-2 km), and the dotted one denotes variance over 2-4 km. The definition of the BL height is provided in Figure 3. Panels (c, d) are the vertical distribution of the LMSE variance $\overline{(h')^2}$ and the moisture variance $\overline{(L_v q'_v - L_f q'_{ice})^2}$ in the troposphere, respectively. The overbar means the horizontal average.

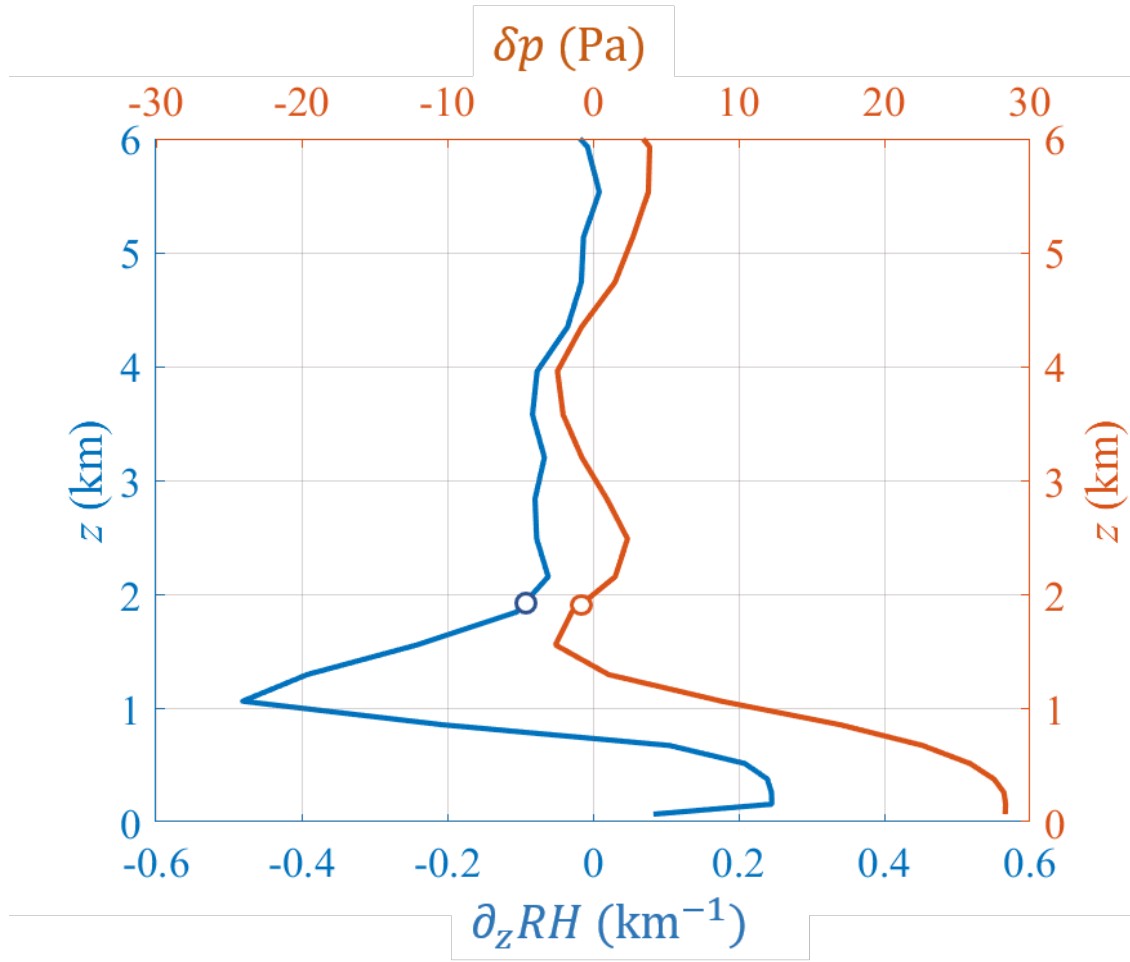

**Figure 3.** Diagnosis for the BL height. The blue line plots the vertical gradient of domain-mean relative humidity $(\partial_z RH)$ averaged over days 50-150. Following Yang (2018a), we define the BL height as the altitude where $\partial_z RH$ first exceeds the threshold of $-0.1$ km$^{-1}$ over the level of the minimum gradient. Therefore, the BL height is about 2 km in our simulation and is labeled by the blue circle. The orange line shows the pressure difference between dry and moist centers $(\delta p)$ at the equilibrium stage. The orange circle labels $\delta p$ at the BL top, and the pressure difference is reasonably small above this level. This suggests that 2 km is also a dynamical BL height where turbulence is well confined below this level.

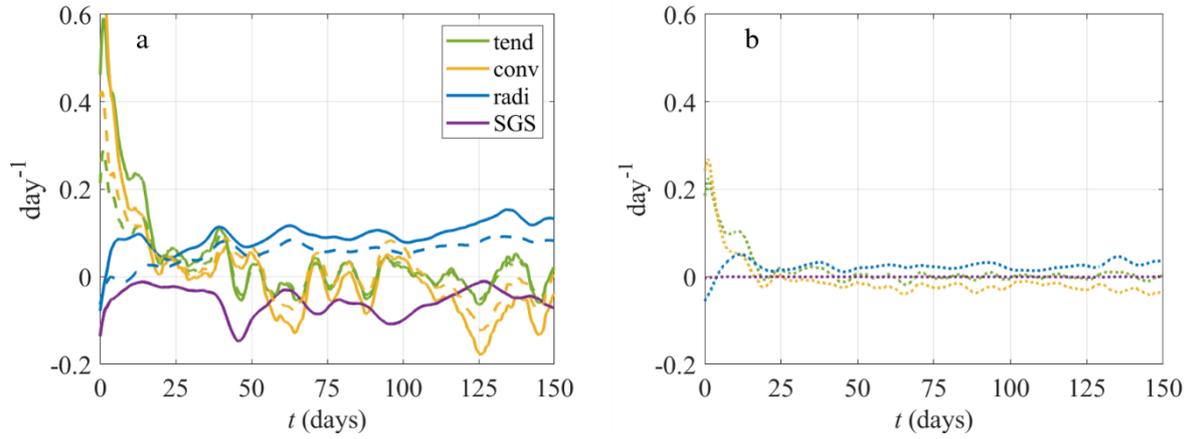

**Figure 4.** The normalized VI-LMSE variance diagnosis over (a) the column, the BL and (b) the lower FT in the CRM simulation [see Eq. (17)]. Solid lines represent the budget integrated over the column ($z_1 = 0, z_2 = z_t$), dashed lines represent that over the BL ($z_1 = 0, z_2 = z_{BL}$), and dotted lines correspond to that over 2-4 km. The dashed curve for SGS is almost identical to the solid curve. Time derivative of the VI-LMSE variance is in green (tend), adiabatic variance production is in yellow (conv), radiative production is in dark blue (radi), and SGS production is in purple (SGS).

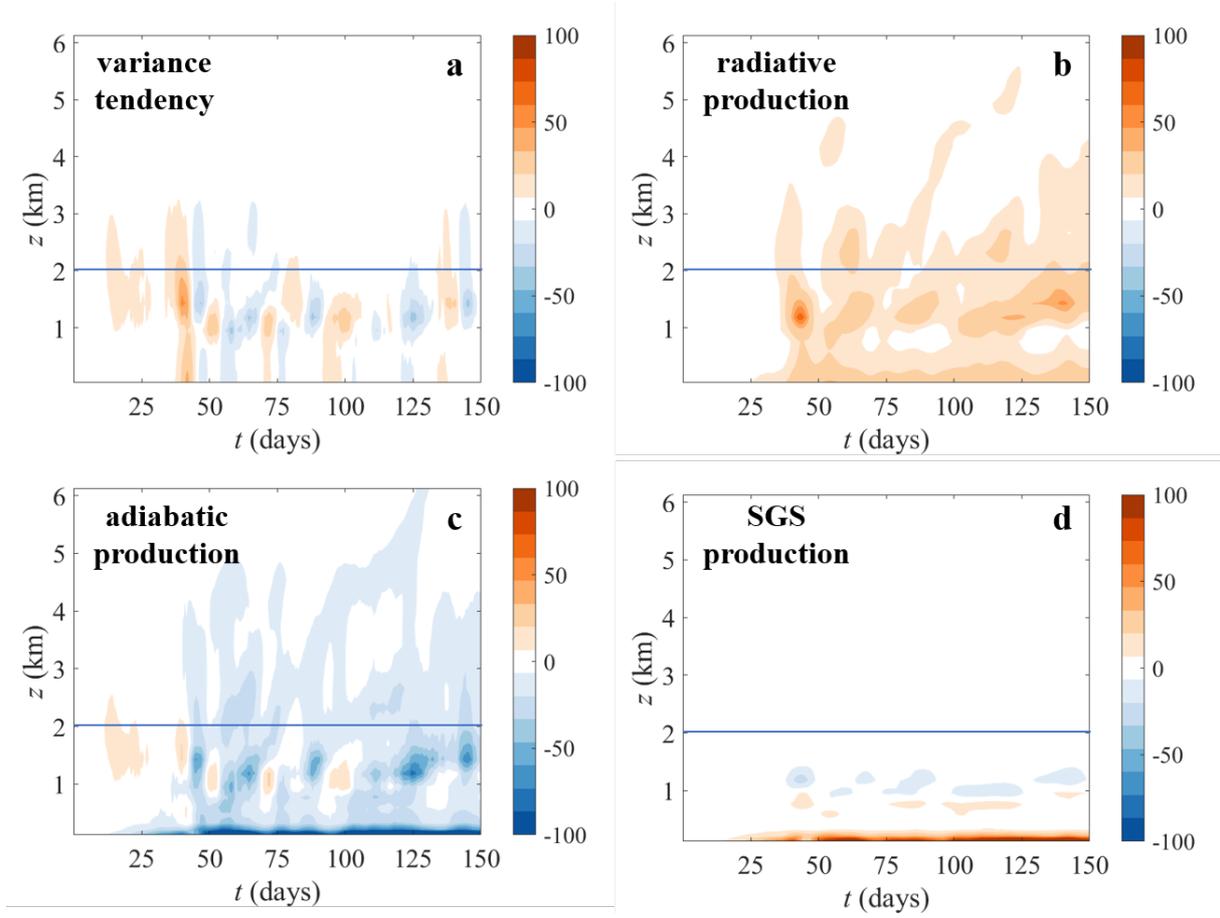

**Figure 5.** The diagnostic results of VR-LMSE variance budget in the CRM simulation [see Eq. (14)]. (a) Variance tendency $\partial_t \frac{1}{2}\overline{(\rho_0 h')^2}$, (b) radiative production $\rho_0^2 \overline{h' Q'_{rad}}$, (c) adiabatic production $\rho_0^2 \overline{h' \left[ -\frac{1}{\rho_0} \nabla \cdot (\rho_0 \vec{v} h)' \right]}$, and (d) SGS production $\rho_0^2 \overline{h' Q'_{sgs}}$. The units are $J^2 m^{-6} s^{-1}$. The lowest model level ($z = 37.5$ m) is excluded in (c) and (d).

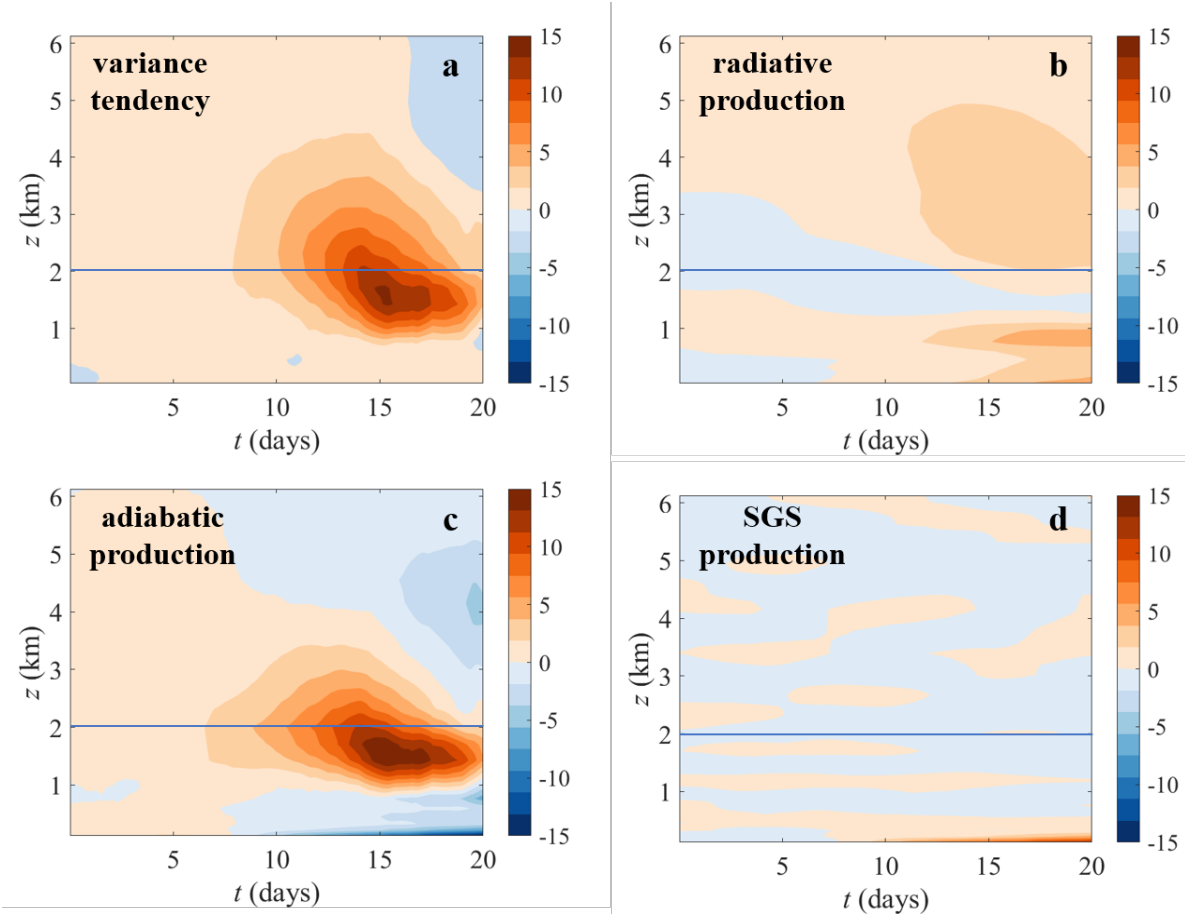

**Figure 6.** Diagnostics of the VR-LMSE variance budget in the first 20 days [see Eq. (14)]. (a) Variance tendency $\partial_t \frac{1}{2}\overline{(\rho_0 h')^2}$ and variance production by (b) radiation $\rho_0^2 \overline{h' Q'_{rad}}$, (c) convergence $\rho_0^2 \overline{h' \left[ -\frac{1}{\rho_0} \nabla \cdot (\bar{\rho}\vec{v}h)' \right]}$, and (d) SGS processes $\rho_0^2 \overline{h' Q'_{sgs}}$ above the near-surface layer. Their units are $J^2 m^{-6} s^{-1}$. The blue lines label the BL top. The lowest level is excluded in (c) and (d).

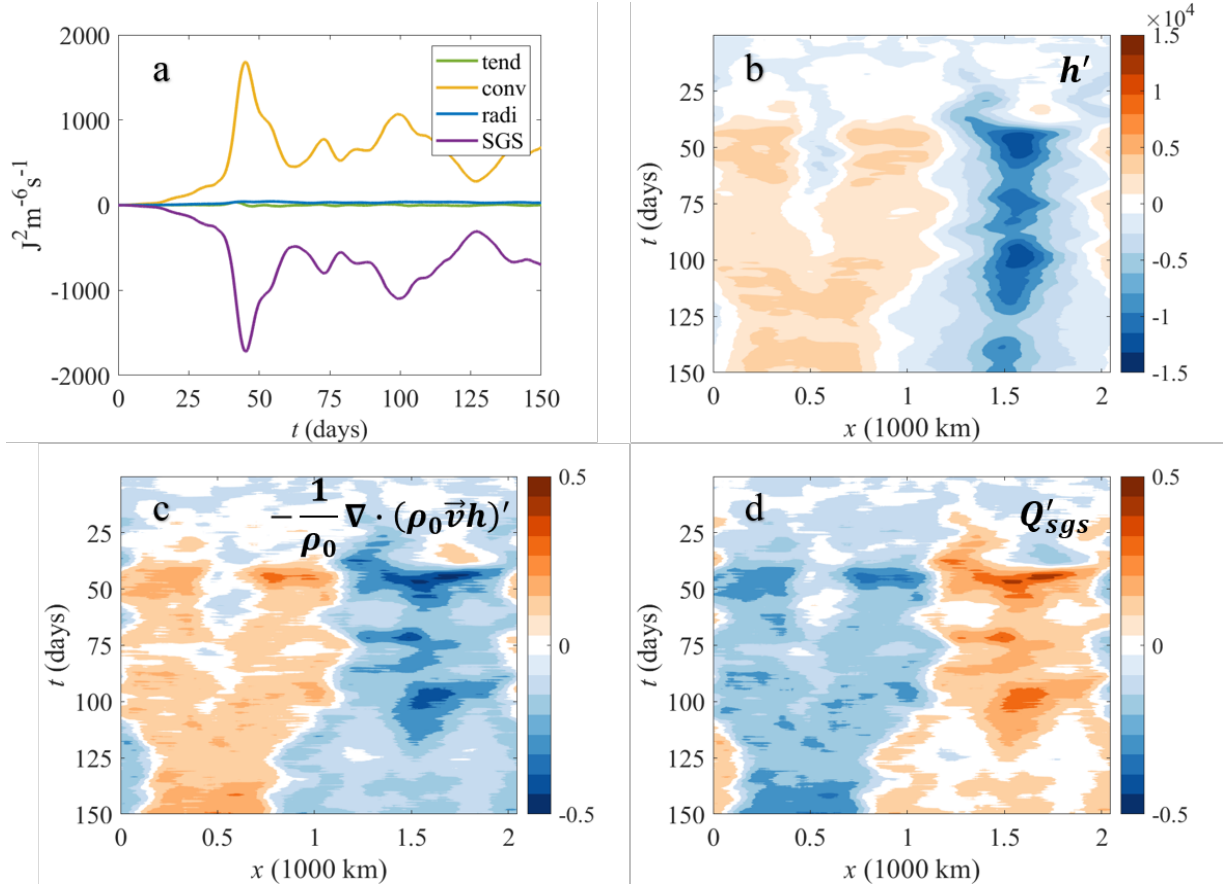

**Figure 7.** (a) The VR-LMSE variance budget at the near-surface layer [see Eq. (14)], and Hovmöller diagrams of (b) near-surface MSE anomalies ($h'$, unit: J kg$^{-1}$) and MSE source anomalies due to (c) near-surface MSE flux convergence ($-\frac{1}{\rho_0}\nabla \cdot (\rho_0 \vec{v} h)'$, unit: W kg$^{-1}$) and (d) near-surface SGS processes ($Q'_{sgs}$, unit: W kg$^{-1}$), respectively. Here, the near-surface layer corresponds to the lowest model level at $z$ = 37.5 m.

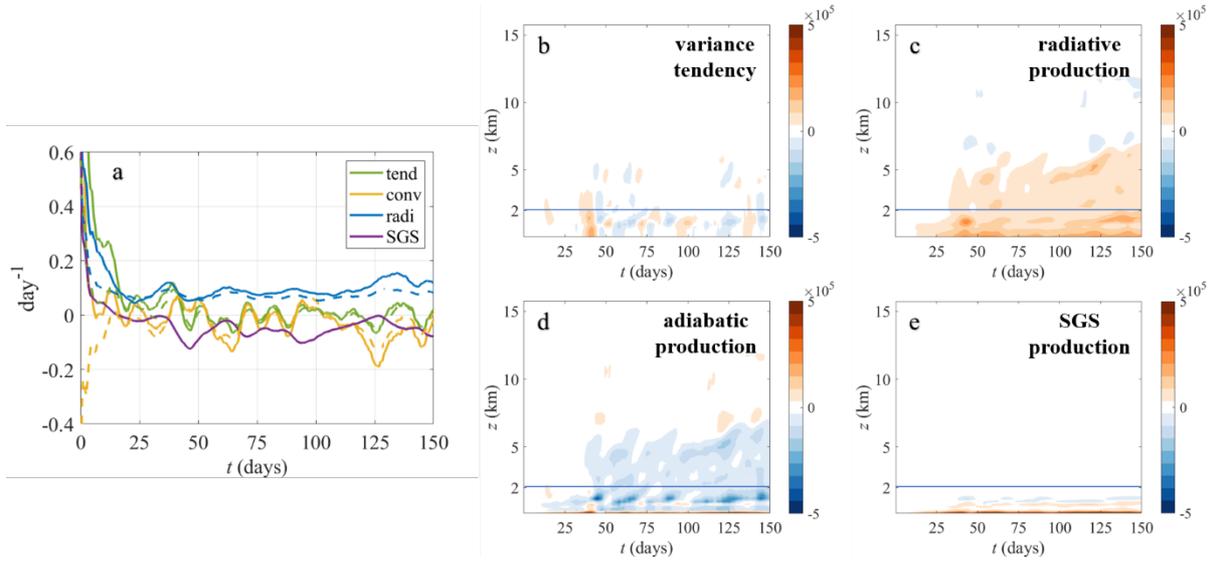

**Figure 8.** The diagnostic results of the GMSE variance framework. (a) The VI-GMSE variance budget, normalized by $\overline{\frac{1}{2}\left(\widehat{h'}\right)^2}$. Solid lines are budgets integrated over the column (from surface to the model top). Dashed lines are budgets integrated over the BL. (b)-(d) The VR-GMSE variance diagnosis [Eq.(18)]. The blue lines indicate the height of the BL. This set of figures shows that the GMSE variance framework also confirms the dominant role of the BL processes on the development of convective self-aggregation after day 15.

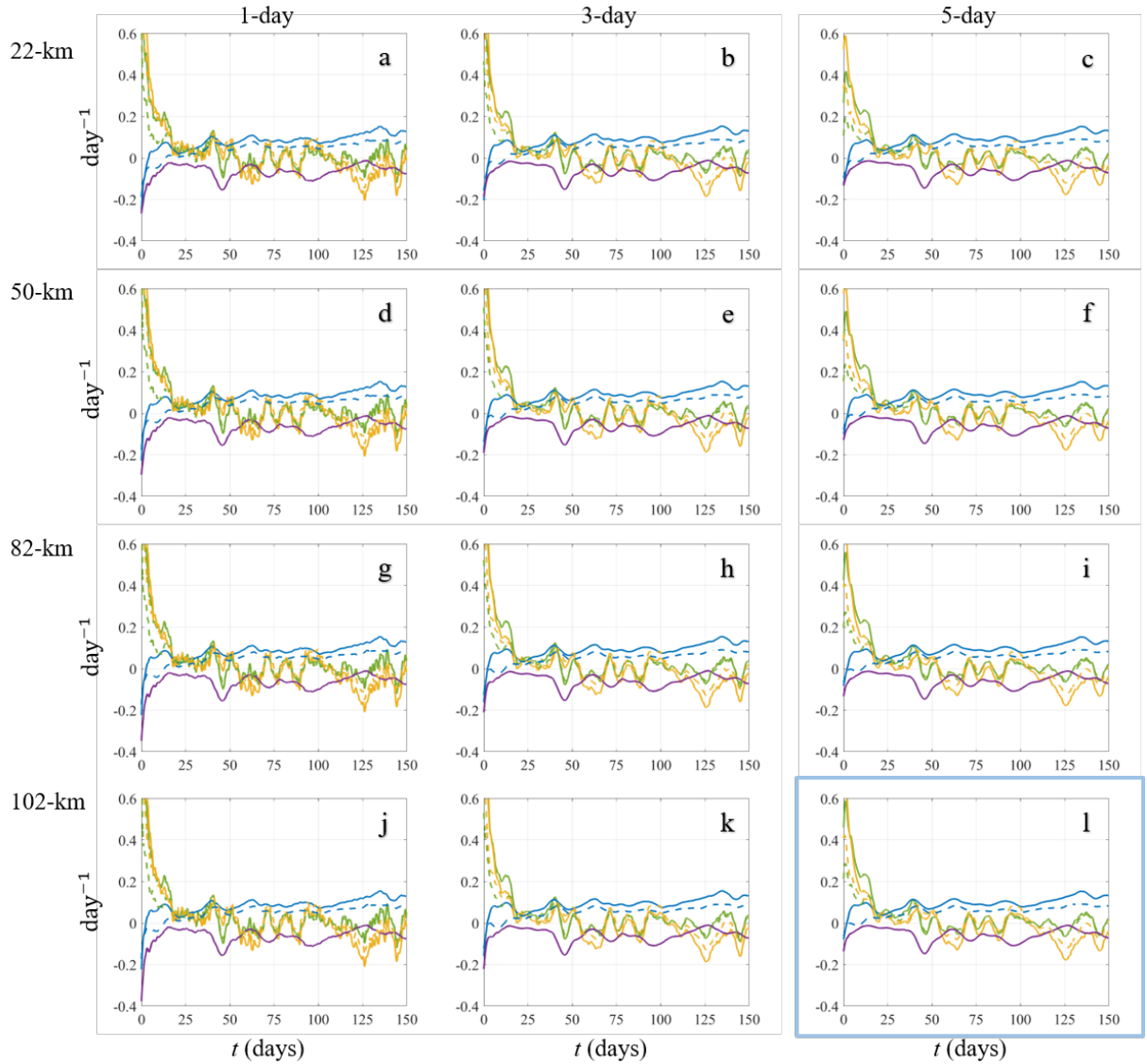

**Figure A1.** Testing the sensitivity of the LMSE variance framework to choices of smoothing window widths. The top to bottom rows correspond to four spatial smoothing windows varying from 22 km to 102 km. The left to right columns are for three temporal smoothing windows: 1 day, 3 days and 5 days. We use the same line colors as Figure 4 to represent different processes. The blue box marks the smoothing windows we use in Figure 4. The LMSE variance diagnostic results are robust to different windows.

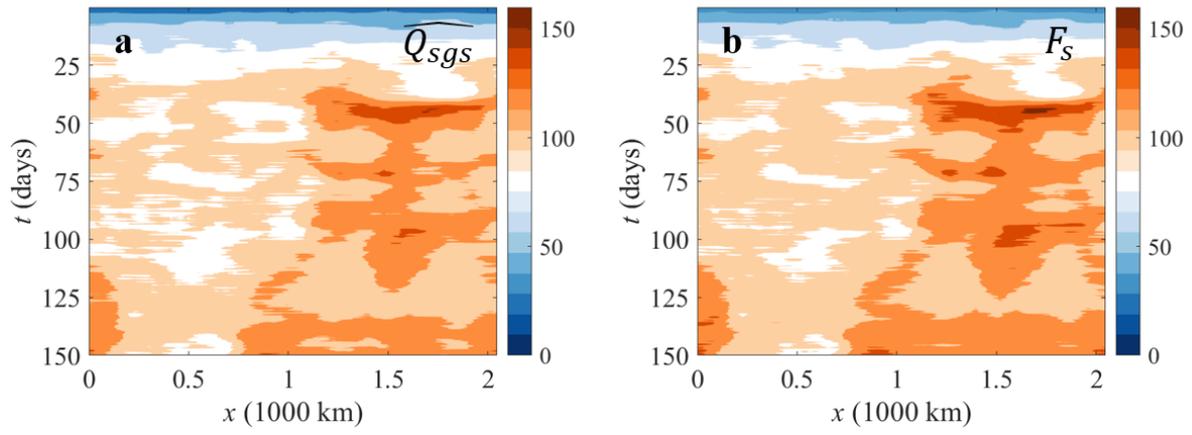

**Figure A2.** (a) The vertically integrated SGS tendencies and (b) the surface enthalpy fluxes. Their units are W m$^{-2}$. Data has been smoothed with 5-day and 102-km smoothing averages.

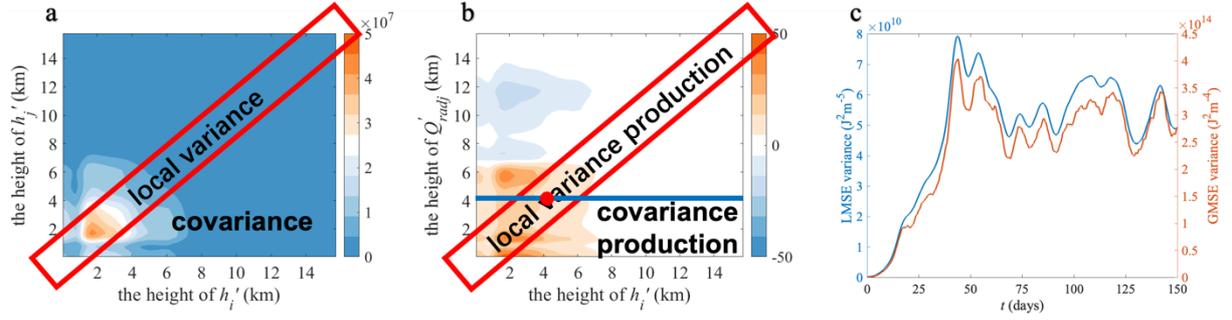

**Figure A3.** Difference in the LMSE and the GMSE variances and their radiative productions in the CRM simulation. Panels (a, b) are matrices of $h'_i \cdot h'_j$ and $Q'_{radj} \cdot h'_i$ at the stable state, respectively (averaged over the last 30 days and in x direction). Here, subscripts $i$ and $j$ represent vertical levels. The local MSE variance in Eq. (10) and the local variance production in Eq. (11) correspond to the diagonal components in the red boxes, where $i = j$. Similarly, the covariance in Eq. (10) and the covariance production in Eq. (11) correspond to the off-diagonal components outside the red boxes, where $i \neq j$. The horizontal blue line represents the variance production at 4 km diagnosed by the VR-GMSE variance budget, while the red dot represents the local variance production diagnosed by the VR-LMSE variance budget. (c) The evolution of the VI-LMSE variance (blue) and the VI-GMSE variance (orange). Both are integrated from the surface to the model top.

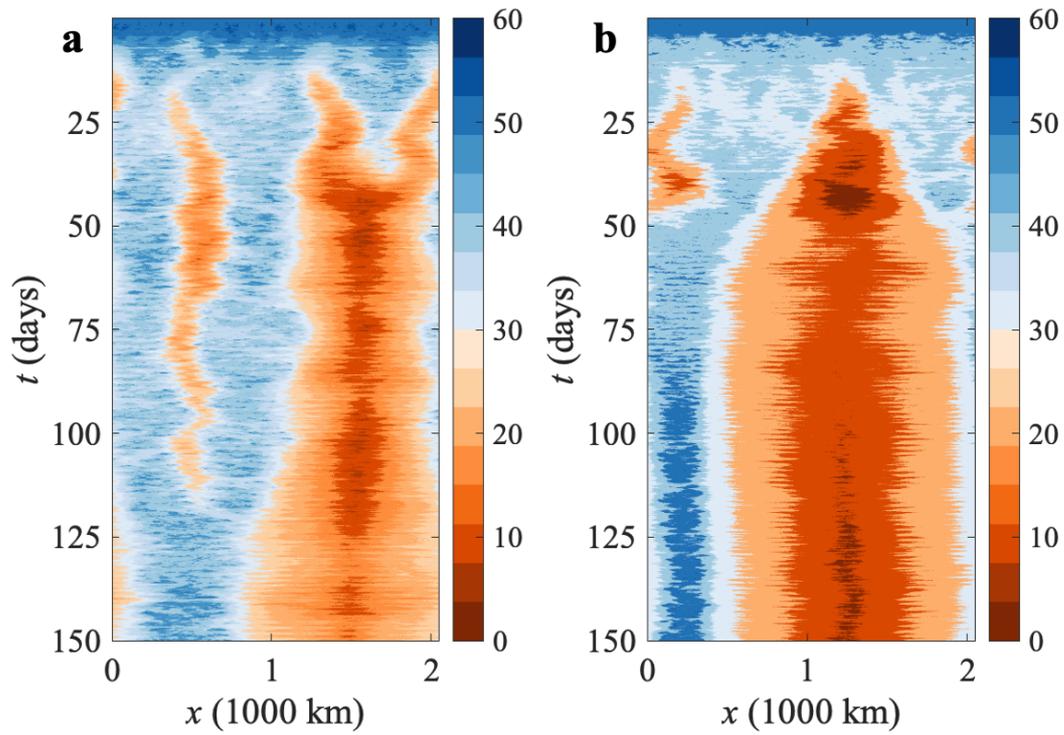

Figure A4. Hovmöller diagrams of the PW in (a) the control simulation and (b) the mechanism-denial simulation. In the mechanism-denial simulation, we homogenize surface fluxes over the domain at each time step. The dry patches in panel (b) form at around day 15, similar to the control simulation.